\documentclass[conference,a4paper]{IEEEtran}
\usepackage{enumerate}
\usepackage{cite}
\usepackage{amscd}
\usepackage{dsfont}
\usepackage{latexsym}
\usepackage{array}

\usepackage{tikz}\usetikzlibrary{calc}
\usepackage{tabularx}
\usepackage{epsfig}
\usepackage{graphicx,subfigure}
\usepackage{amsmath,mathrsfs}
\usepackage{amsthm}
\usepackage{amssymb}  
\usepackage{amsbsy}
\usepackage{color}
\usepackage{stfloats}
\usepackage{algorithm}
\usepackage{algpseudocode}
\usepackage{bm}
\usepackage{pifont}
\usepackage{accents}
\usepackage{multirow}
\usepackage{colortbl}

\newtheorem{lem}{Lemma}

\newtheorem{deft}{Definition}

\theoremstyle{definition}
\newtheorem{rem}{Remark}

\ifCLASSINFOpdf
\else
\fi
\begin{document}
%
\title{Set Reconciliation for Blockchains with Slepian-Wolf Coding: Deletion Polar Codes}

\author{
\IEEEauthorblockN{Ling Liu}
\IEEEauthorblockA{Shenzhen University\\
Guangdong, China\\
Email: liulingcs@szu.edu.cn}
\and
\IEEEauthorblockN{Shengli Zhang}
\IEEEauthorblockA{Shenzhen University\\
Guangdong, China\\
Email: zsl@szu.edu.cn}

\and

\IEEEauthorblockN{Cong Ling}
\IEEEauthorblockA{Imperial College London\\
London, UK\\
Email: cling@ieee.org}

}


%


\maketitle

\begin{abstract}
In this paper, we propose a polar coding based scheme for set reconciliation between two network nodes. The system is modeled as a well-known Slepian-Wolf setting induced by a fixed number of deletions. The set reconciliation process is divided into two phases: 1) a deletion polar code is employed to help one node to identify the possible deletion indices, which may be larger than the number of genuine deletions; 2) a lossless compression polar code is then designed to feedback those indices with minimum overhead. Our scheme can be viewed as a generalization of polar codes to some emerging network-based applications such as the package synchronization in blockchains. Some connections with the existing schemes based on the invertible Bloom lookup tables (IBLTs) and network coding are also observed and briefly discussed.
\end{abstract}


%
\IEEEpeerreviewmaketitle

\section{Introduction}
The increasing scale of data in toady's cloud, network environment and other distributed systems requires much higher bandwidth consumption for the purpose of package synchronization among replicas, even if there are slight data differences. Efficient synchronization protocols or algorithms are crucial for emerging network-based applications such as blockchains, which keeps propagating fresh transactions and mined blocks among different nodes. Guaranteeing those transactions and blocks consistent and synchronized is important for both commercial and secure purposes. Great effort has been made in this direction during the recent years. Many excellent synchronization methods and protocols have been developed, and most of them use the popular data structures such as Bloom filters \cite{BloomISIT2012} and invertible Bloom lookup tables (IBLTs) \cite{IBLT2011} for set reconciliation. As a seminal probabilistic data structure, Bloom filters can efficiently check whether an element is a member of a set, with high successful probability. With a richer structure, the IBLTs can not only check the set difference, but also recover the missing items. An interesting connection between IBLTs and classical graph-based erasure codes has been observed, as they both rely on ``checksum" bits and use similar onion-peering decoding algorithms. This connection makes existing erasure and error correction codes good candidates for the set reconciliation problem.

Specifically, we model the set reconciliation problem as a modified  Slepian-Wolf setting, with fixed number of deletions. This work investigates the potential of polar codes in this direction. Being considered as a major breakthrough in coding theory, polar codes \cite{arikan2009channel} are the first kind of theoretically provable capacity achieving codes for binary-input memoryless symmetric channels (BMSCs). The novel channel polarization technique enables polar codes to achieve channel capacity by successive cancellation (SC) decoding with low complexity. More sophisticated decoding algorithms such as belief propagation (BP) decoding \cite{Eslami2012BPfinite}, successive cancellation list (SCL) decoding \cite{ListPolar} and successive cancellation stack (SCS) decoding \cite{NiuKaiSCS2013} have been proposed later. The versatility of polar codes has then been witnessed at other scenarios including asymmetric channels \cite{aspolarcodes}, wiretap channels \cite{polarsecrecy}, broadcast channels \cite{polarbroadcast}, multiple access channels \cite{AbbeMac} and even quantum channels \cite{PolarQuantum}. More recently, polar codes found their application in channels with deletions \cite{PolarDeletionsRKLiu,PolarDeletionVTan,PolarDeletionTal}. The so-called deletion polar codes will be a key ingredient of our polar coding based set reconciliation protocol.

Another ingredient of our protocol is polar coding for lossless compression. Besides channel coding, polar codes can be also extended to source coding, for both lossless \cite{cronie2010lossless} and lossy compression \cite{KoradaSource}. The corresponding source polarization technique was introduced to solve the Slepian-Wolf problems with perfect synchronization over symbols \cite{polarsource}. In our case of set reconciliation, where synchronization is not available, a deletion polar code is first designed to aid one peer to locate the possible deletions, which helps to obtain some relaxed synchronization information, and then the possible deletion indices are losslessly compressed and returned. The optimality of polar codes for lossless compression provides us negligible overhead for this step.

The rest of the paper is organized as follows: Section II presents a brief introduction of our system model and the overview scheme. A bit-wise Slepian-Wolf problem with fixed number of deletions is then defined between two peers. The details for identifying the possible deletion locations are presented in Section III, where we employ deletion polar codes to align one particular column of data for the two peers. Then, we design a deletion detection algorithm to locate the potential deletions based on the aligned bit stream. We also show that the amount of potential deletions is roughly three times of the genuine deletion number after one round of data alignment, which is also verified by numerical simulation. With the assistant of the potential deletions, the system is converted to a Slepian-Wolf problem with erasures in Section IV. By approximating the occurrence of the potential deletions as a Bernoulli source model, a polar coding based lossless compression scheme is utilized to return the missing indices. Finally, the paper is concluded in Section V.

All random variables are denoted by capital letters. For a set $\mathcal{I}$, $\mathcal{I}^c$ denotes its complement, and $|\mathcal{I}|$ represents its cardinality. Following the notation of \cite{arikan2009channel}, we use $X_1^N$ as a short hand of a row vector $(X_1, ..., X_N)$. Let $[N]$ denote the set of all integers from 1 to $N$. For a subset $\mathcal{I} \subset [N]$, $X^{\mathcal{I}}$ represents the subsequence of $X_1^N$ with indices in $\mathcal{I}$.

\section{System Model and Overview Scheme}
\begin{figure}[ht]
    \centering
    \includegraphics[width=9cm]{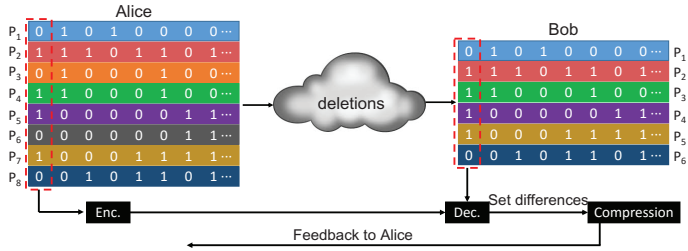}
    \caption{The system model of the set reconciliation problem with deletions.}
    \label{fig:DelModel}
\end{figure}

The graphical settings of the set reconciliation problem are depicted in Fig. \ref{fig:DelModel}. A set of package data is shared between the two peers Alice and Bob through a network, where Alice is the central node and she is assumed to have a complete data backup, while Bob has an incomplete backup with a certain amount of missing packages. We notice that the noise is modeled as deletions instead of erasures since the indices of missing packages are unknown on Bob's side. A package is represented as a binary row vector with length $L$ in Fig. 1. We also assume there are $N=2^n$ packages on Alice's side for the convenience of the following polar coding. Please note that all the packages follow a chronological ordering, which can be obtained from their corresponding content. This assumption is natural because in many network applications the package data contains a precise version of its generation time. Particularly, such a package in blockchains may represent a transaction record, which always contains its accurate time information. The unstable network conditions lead to several package deletions on Bob's side. In Fig. 1, the third and the sixth packages are deleted from Alice's perspective. Thanks to the time information, Bob can still order the remaining packages chronologically.

\begin{figure}[ht]
    \centering
    \includegraphics[width=9cm]{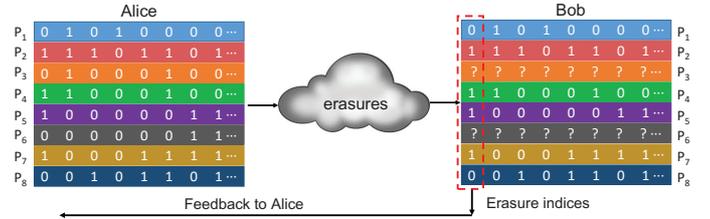}
    \vspace{-1em}
    \caption{The system model of the set reconciliation problem with erasures.}
    \label{fig:EraModel}
\end{figure}

\begin{rem}
We note that when perfect package synchronization is available, the connection between the two nodes can be modeled as a channel with a certain amount of erasures instead of deletions, as shown in Fig. \ref{fig:EraModel}. The set reconciliation task at this scenario is much simpler as Bob can directly identify the indices of the missing packages from his local data. However, for the set reconciliation problem with deletions, more effort is required to obtain those indices, and we shall see that polar codes are promising in addressing this issue with very small data overhead.
\end{rem}

\begin{figure}[ht]
    \centering
    \includegraphics[width=7cm]{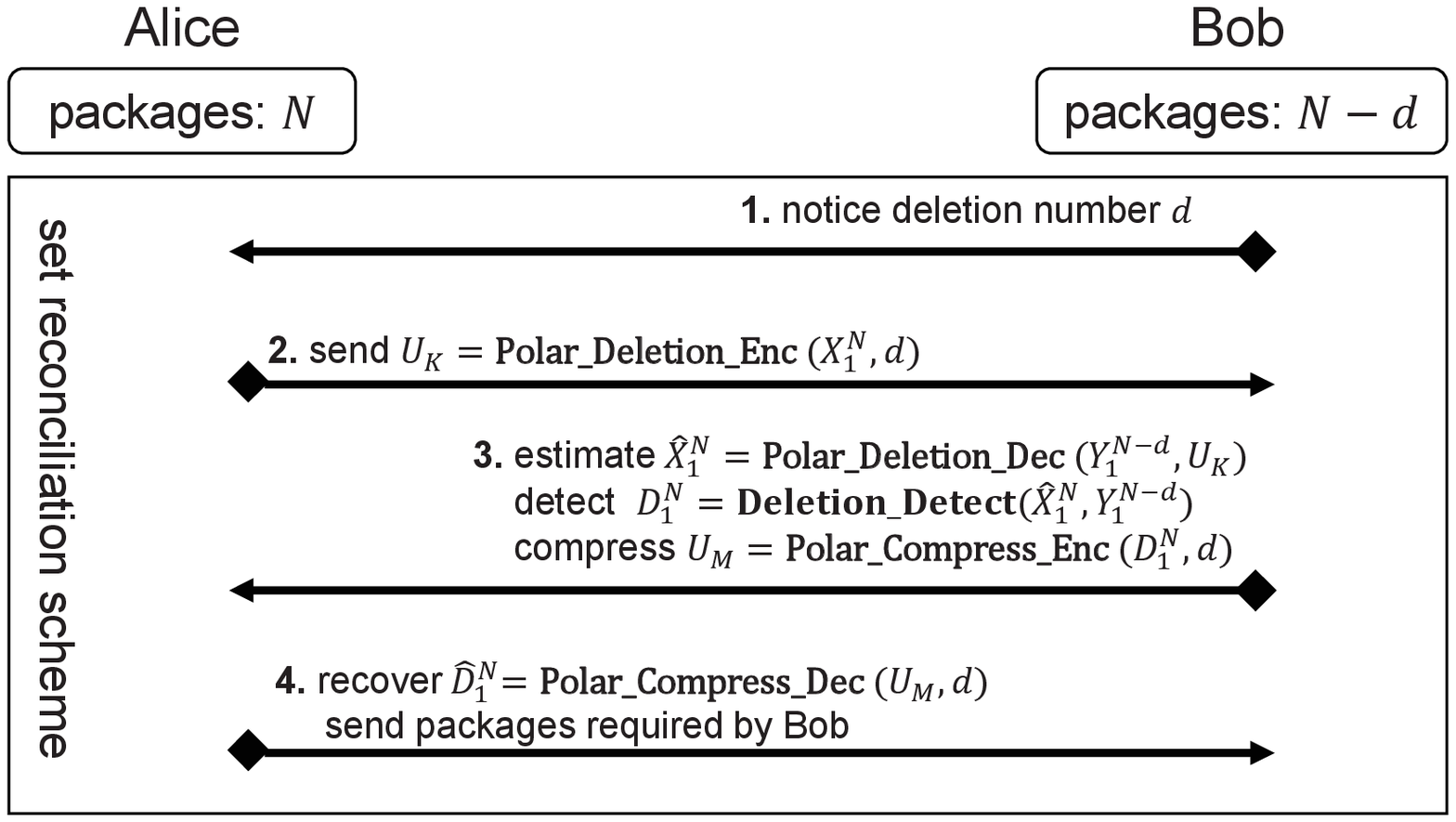}
    \caption{A high-level description of the proposed set reconciliation scheme using polar codes.}
    \label{fig:DelScheme}
\end{figure}

Our proposed scheme can be summarized as in Fig. \ref{fig:DelScheme}. It starts when Alice collects $N$ packages, and $N$ is a preset number known to both Alice and Bob. Bob then counts the amount of his local packages, and informs Alice the number of deletions $d$. After knowing $d$, Alice and Bob pick one specific column on both sides to detect the location of deletions. Since all bits of package data are assumed to be uniformly random, the first column (see Fig. 1) is picked without loss of generality. Alice then encodes her first column data $X_1^N$ into $K$ bits $U_K$
using a deletion polar encoding function \textbf{Polar\_Deletion\_Enc}, and sends $U_K$ to Bob. With the assistance of the received $K$ bits, Bob tries to recover the estimation $\hat{X}_1^N$ of $X_1^N$ from his local data $Y_1^{N-d}$ with the decoding function \textbf{Polar\_Deletion\_Dec}. The two data columns are then aligned by the function \textbf{Deletion\_Detect} to identify the potential deletion positions, or equivalently the set difference, which can be expressed by a binary sequence $D_1^N$ with ``1'' denoting the potential deletion. Next, $D_1^N$ is compressed to $U_M$ ($M \leq N$) by the function \textbf{Polar\_Compress\_Enc} to further reduce the overhead. Alice finally obtains $D_1^N$ from $U_M$ using \textbf{Polar\_Compress\_Dec} and sends the required packages. This protocol can be viewed as a solution to the Slepian-Wolf problem for two joint binary symmetric sources (BSSs) with deletions.

For the example in Fig. 1, we have $N=8$ and $d=2$. By aligning the two column vectors $X_1^8=[01011010]$ and $Y_1^6=[011110]$, Bob knows the potential deletion positions are 3 and 6, and the set difference is described by $D_1^8=[00100100]$ consequently. We note that it is not always the case that Bob obtains the exact deletion positions. As we shall see, the number of potential deletions generally gets larger than $d$. However, since $d$ is relatively smaller compared with $N$, the resulted sequence $D_1^N$ is quite biased, which explains the motivation of the further lossless compression process.

\section{Polar codes for Slepian-Wolf problem with deletions}\label{sec:Algebfading}
\subsection{Polar Codes for Deletions}
Let $W$ be a BMSC with input alphabet $X$ and output alphabet $Y$. Given the capacity $C$ of $W$ and a rate $R<C$, the information bits of a polar code with block length $N=2^m$ are indexed by a set of $\lfloor RN \rfloor$ rows of the generator matrix $G_N=B\cdot F^{\otimes n}$, where $F=\left[\begin{smallmatrix}1&0\\1&1\end{smallmatrix}\right]$, $\otimes$ denotes the Kronecker product, and $B$ is the bit-reverse permutation matrix. The matrix $G_N$ combines $N$ identical copies of $W$ to $W_N$. Then this combination can be successively split into $N$ binary memoryless symmetric subchannels, denoted by $W_{N}^{(i)}$ with $1 \leq i \leq N$. By channel polarization, the fraction of good (roughly error-free) subchannels is about $C$ as $n\rightarrow \infty$. Therefore, to achieve the capacity, information bits should be sent over those good subchannels and the rest are fed with frozen bits which are known before transmission. The indices of good subchannels are generally identified according to their associated Bhattacharyya Parameters.
\begin{deft}\label{deft:symZ}
Given a BMSC $\tilde{W}$ with transition probability $P_{Y|X}$, the Bhattacharyya parameter $Z\in[0,1]$ is defined as
\begin{eqnarray}
Z(W)=Z(X|Y)&\triangleq\sum\limits_{y} \sqrt{P_{Y|X}(y|0)P_{Y|X}(y|1)}.
\end{eqnarray}
\end{deft}

Based on the Bhattacharyya parameter, the information set $\mathcal{I}$ is defined as $\{i:Z(W_{N}^{(i)})\leq 2^{-N^{\beta}}\}$ for any $0<\beta <\frac{1}{2}$, and the frozen set $\mathcal{F}$ is the complement of $\mathcal{I}$. Let $P_B$ denote the block error probability of a polar code under the SC decoding. It can be upper-bounded as $P_{B}\leq\Sigma_{i\in \mathcal{I}}Z((W_{N}^{(i)})$. Efficient algorithms to evaluate the Bhattacharyya parameter of subchannels for general BMSCs were presented in \cite{IdoConstruct,PolarConstru,mori2009performance}.

However, when $W$ is a deletion channel with fixed deletion numbers $d$, which is no longer memoryless, the design of polar codes becomes more complicated. In fact, the polarization phenomenon can be generalized to the memory cases \cite{PolarMemorySasoglu,PolarMemoryTal}. Particularly, for a deletion channel with $d$ deletions, although its channel capacity is still unknown, the trend of polarization has been well observed \cite{PolarDeletionsRKLiu} and further proved \cite{PolarDeletionTal}. Moreover, a practical modified SC decoding algorithm was proposed in \cite{PolarDeletionsRKLiu}, which has a complexity of roughly $O(d^2 N \log N)$. Compared with the previous work on deletion channels \cite{PolarDeletionVTan}, which exhaustively searches all possible deletion patterns and then perform SC decoding, \cite{PolarDeletionsRKLiu} suggests to use a state triple $(d_1, d_2, d_3)$ to label every consecutive sequence, where $d_1$, $d_2$ and $d_3$ denote the number of deletions before, within and after the sequence, respectively. For the modified SC decoding, a parent node with a certain state triple corresponds to two kid nodes with their state triples being complementally coupled. In this work, we adopt the decoding scheme in \cite{PolarDeletionsRKLiu} to solve the Slepian-Wolf problem with deletions.

\begin{figure}[ht]
    \centering
    \includegraphics[width=7cm]{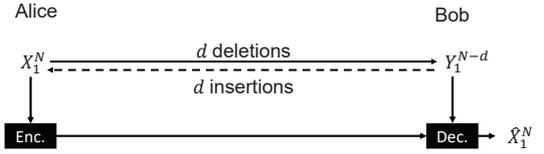}
    \vspace{-1em}
    \caption{The Slepian-Wolf problem with deletions.}
    \label{fig:SWDel}
\end{figure}

The Slepian-Wolf coding problem between two joint BSSs with deletions is lifted from the system model, as shown in Fig. \ref{fig:SWDel}, where $X_1^N \in \{0,1\}^{N}$ denotes $N$ i.i.d. random variables with uniform distribution and $Y_1^{N-d} \in \{0,1\}^{N-d}$ represents a noisy copy of $X_1^N$ with $d$ random deletions. A polar code constructed for channels with deletions can be easily adopted here to solve this problem. Let $U_1^N=X_1^N G_N$ denote the encoded bits after the polar transform. In order to reconstruct $X_1^N$ or equivalently $U_1^N$ on Bob's side, the decoder needs to know the unreliable bits $U^{\mathcal{F}_d}$ in $U_1^N$. For any given $0<\delta<1$, the set $\mathcal{F}_d$ is defined as
\begin{eqnarray}
\mathcal{F}_d\triangleq \{i \in [N]: P_d^{SC}(U_i|U_1^{i-1},Y_1^{N-d}) > \delta\},
\end{eqnarray}
where $P_d^{SC}(U_i|U_1^{i-1},Y_1^{N-d})$ denotes the error probability of the $i$-th subchannel by the modified SC decoding method \cite{PolarDeletionsRKLiu}.

\begin{figure}[ht]
    \centering
    \includegraphics[width=8cm]{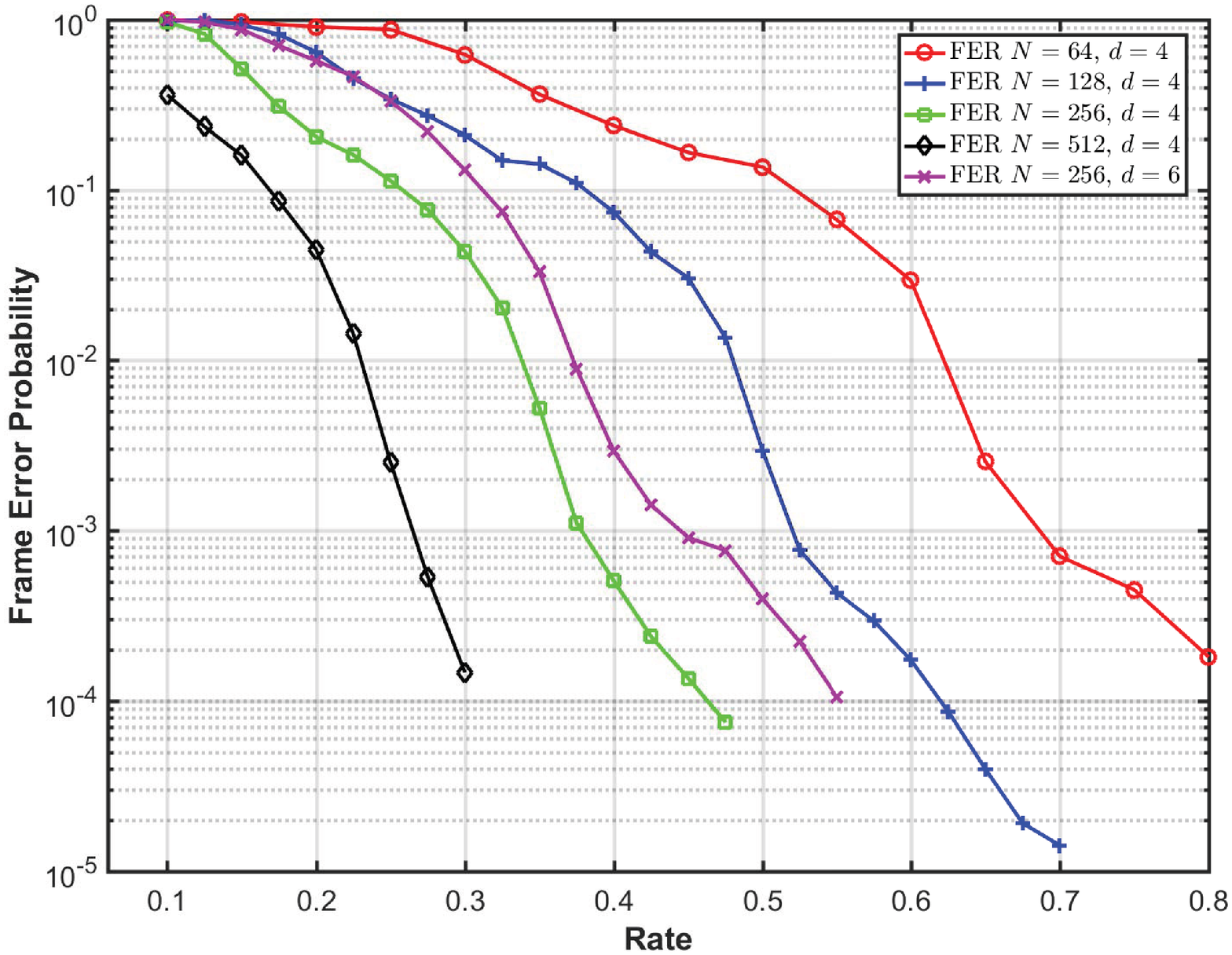}
    \caption{The performance of polar codes for Slepian-Wolf problems with deletions.}
    \label{fig:SW_Del_FER}
\end{figure}

Unfortunately, the existing evaluation methods of $Z(W_N^{(i)})$ for memoryless channels cannot be used to evaluate $P_d^{SC}(U_i|U_1^{i-1},Y_1^{N-d})$ for deletion channels. In practice, one can use the Monte Carlo method to estimate $P_d^{SC}(U_i|U_1^{i-1},Y_1^{N-d})$, which is only determined by $N$ and $d$. Therefore, the estimation can be performed off-line and pre-shared between Alice and Bob. After that, $P_d^{SC}(U_i|U_1^{i-1},Y_1^{N-d})$ is sorted in descend order and the first $K$ indices form the set $\mathcal{F}_d$, namely $|\mathcal{F}_d|=K$ and the coding rate $R=\frac{K}{N}$. Note that we use $U_K$ to represent $U^{\mathcal{F}_d}$ for convenience. Once receiving the bits $U_K$, Bob implements the decoding algorithm to recover the remaining bits in $U_1^N$, treating $Y_1^{N-d}$ as the channel output of the deletion channel. The performance of polar codes of varying rates for different $N$ and $d$ is illustrated in Fig. \ref{fig:SW_Del_FER}. It can be seen that the performance gets better when $N$ increases or $d$ decreases.

\begin{rem}\label{rk:DeletionRvs}
As shown in Fig. \ref{fig:SWDel}, our Slepian-wolf coding scheme can also be performed reversely from Bob to Alice with the encoder and the decoder being swapped. In this case, we view the channel from Bob to Alice as a channel with $d$ insertion (see the dash line). The design of polar codes for insertion channels follows a similar idea as that for deletion channels \cite{PolarDeletionsRKLiu}. In this work, we prefer the setting of deletion channels to make it consistent with \cite{PolarDeletionsRKLiu}.
\end{rem}

\begin{rem}\label{rk:DeletionCapacity}
Generally speaking, the set difference is relatively much smaller than the size of the whole package set, i.e., $\frac{d}{N}$ is a small number less than $0.5$. In this case, the capacity of the channel with $d$ deletions is lower-bounded by $1-h_2(1-\frac{d}{N})$ \cite{DeletionCapacity}, where $h_2(\cdot)$ is the binary entropy function. By the extremal property of polarization, the coding rate for our Slepian-Wolf coding scheme can be upper-bounded by $h_2(1-\frac{d}{N})$, when $N$ is sufficiently large.
\end{rem}

\begin{rem}\label{rk:CheckSum}
We note that the rationale behind our polar-coding based set reconciliation scheme still matches that of some existing schemes (e.g. \cite{Graphene2019}) using Bloom filters and IBLTs in the sense that some ``checksum" bits of local data are sent to assist the other peer for reconstruction and then set comparison. More explicitly, $U_K$ is the ``checksum" of $X_1^N$  according to the matrix $G_{\mathcal{F}_d}$, which is a submatrix of $G_N$ with column indices in $\mathcal{F}_d$. The difference is that the ``checksum" bits are  generated from a single bit of each package instead of the entire bits within it, which makes the complexity of our scheme uncorrelated to the size of each package.
\end{rem}

\begin{table}[H]
\begin{center}
\caption{An example of the admissible table}
\label{tab:Adm}
\begin{tabular}{|c||c|c|c|}\hline
& \multicolumn{3}{c|}{deletion state $(d_2,d_1)$}\\\hline
$i$ & $(d_2=0,d_1=0)$ & $(d_2=0,d_1=1)$ & $(d_2=1,d_1=0)$  \\\hline\hline
$1$  &  1 $\blacksquare\bigstar\blacktriangle$  &  0  &  1     \\\hline
$2$  &  1 $\blacksquare\bigstar\blacktriangle$  &  0  &  1     \\\hline
$3$  &  1 $\blacksquare\bigstar$  &  0  &  1 $\blacktriangle$    \\\hline
$4$  &  1 $\blacksquare$  &  1 $\blacktriangle$  &  1   $\bigstar$  \\\hline
$5$  &  0  &  1 $\bigstar\blacktriangle$  &  1 $\blacksquare$     \\\hline
$6$  &  1  &  1 $\blacksquare\bigstar\blacktriangle$  &  1     \\\hline
$7$  &  0  &  1 $\blacksquare\bigstar\blacktriangle$  &  1     \\\hline
$8$  &  0  &  1 $\blacksquare\bigstar\blacktriangle$  &  1     \\\hline
\end{tabular}
\end{center}
\end{table}

\subsection{Deletion Detection}
After Bob reconstructs $X_1^N$ successively, a detection algorithm is employed to locate the potential deletion positions, based on an admissible table data structure. The admissible table is denoted by an $N \times (2d+1)$ binary matrix $T$. The row index $i$ of $T$ corresponds to the bit index of $[N]$, and its column index $j$ corresponds to a state vector $(d_2, d_1)$, where $d_1$ and $d_2$ represent the number of deletions before and within the $i$-th bit, respectively. We can easily check that $d_1 \geq 0$, $ 0\leq d_2\leq 1$ and $d_1+d_2 \leq d$. Therefore, there are $2d+1$ columns with $(d_2, d_1)=(0, 0), ..., (0,d), (1,0), ..., (1, d-1)$, and we have $j=d_2\times (d+1)+ d_1+1$. The element $T(i,j)$ represents whether the $j$-th state for the $i$-th bit is admissible or not. The pseudo-code of obtaining $T$ is given in Algorithm \ref{Alg:alg1}.

\begin{algorithm}[H]
  \caption{The generation of the admissible table}
  \label{Alg:alg1}
  \begin{algorithmic}[1]
  \State \textbf{Input}: $X_1^N$ (Alice's 1-st column), $Y_1^{N-d}$ (Bob's 1-st column).
  \State \textbf{Output}: $T$ (admissible table)
  \For{$i=1:N$}
    \For{$j=1:2d+1$}
      \If{$j<d+2$}
      \State$d_1=j-1$
        \If{$d_1>i-1$ or $d_1<i-(N-d)$}
         \State $T(i,j)=0$
         \Else
         \State $T(i,j)=(X_i==Y_{i-d_1})$
        \EndIf
        \Else
        \State $d_1=j-d-2$
        \If{$d_1>i-1$ or $d_1<i-(N-d)-1$}
          \State $T(i,j)=0$
        \Else
          \State $T(i,j)=1$
        \EndIf
      \EndIf
    \EndFor
  \EndFor
   \State Return $T$
  \end{algorithmic}
\end{algorithm}

\begin{table}[H]
\begin{center}
\caption{An example of the path-checking table}
\label{tab:Path}
\begin{tabular}{|c||c|c|c|}\hline
& \multicolumn{3}{c|}{patential paths}\\\hline
$i$ & Path $\#1$ & Path $\#2$ & Path $\#3$  \\\hline\hline
$1$  &  1  &  1  &  1     \\\hline
$2$  &  1  &  1  &  1     \\\hline
$3$  &  1  &  1  &  3     \\\hline
$4$  &  1  &  3  &  2     \\\hline
$5$  &  3  &  2  &  2     \\\hline
$6$  &  2  &  2  &  2     \\\hline
$7$  &  2  &  2  &  2     \\\hline
$8$  &  2  &  2  &  2     \\\hline
\end{tabular}
\end{center}
\end{table}

An example of the admissible table when $X_1^8=[1 0 1 1 1 0 0 1]$ and $Y_1^7=[1 0 1 1 0 0 1]$ is shown in Table \ref{tab:Adm}. Based on the admissible table, a path-checking algorithm is developed to detect the potential paths of the state evolution. We notice that a potential path of the deletion state propagation can only go through the $``1"$ elements in the admission table. For example, when $i=1$, there are two available states in the first row of Table \ref{tab:Adm}, i.e., State 1 ($d_2=0, d_1=0$) and State 3 ($d_2=1, d_1=0$). If State 1 is chosen, $d_1$ will not be changed for the next index $i=2$, and there will be two admissible states for $i=2$ as well. However, if State 3 is chosen for $i=1$, $d_1$ will increase to $1$ for $i=2$, and the path prorogation will be terminated since State 2 ($d_2=0, d_1=1$) for the second row is inadmissible. The pseudo-code of checking the potential paths of the state evolution is given in Algorithm \ref{Alg:alg2}, where a path propagation function \textbf{PathProp} is iteratively called. The detail of \textbf{PathProp} is given in Algorithm \ref{Alg:alg3}.

For the admissible table in Table \ref{tab:Adm}, the path-checking table is shown in Table \ref{tab:Path}. The three potential paths correspond to the evolution of deletion state in Table \ref{tab:Adm} are labeled with symbol $\blacksquare$, $\bigstar$ and $\blacktriangle$, respectively. It can be seen that the positions of the potential deletions for each path are the indices of elements larger than $d+1$. In Table \ref{tab:Path}, the deletion may occur for $i=3, 4$ or $5$, namely $\mathcal{D}=\{3,4,5\}$.

Let $\hat{d}$ denote the number of potential deletions after the previously introduced detection algorithm for one data column. From the above example we can see that $\hat{d}\geq d$. Since the potential deletion indices need to be returned to Alice, who may further process these $\hat{d}$ packages. A natural question is that how large $\hat{d}$ is with respect to $d$. The following lemma gives an upper-bound on $\hat{d}$ for one data column alignment.

\begin{algorithm}
  \caption{The path-checking algorithm over the admissible table}
  \label{Alg:alg2}
  \begin{algorithmic}[1]
  \State \textbf{Input}: the admissible table $T$
  \State \textbf{Output}: the path-checking table $P$ and potential deletion indices $\mathcal{D}$
  \State $[N,M]=$ \textbf{sizeof}($T$)
  \State $d=(M-1)/2$
  \State $Cnt=0$
  \State $Path$ = \textbf{zeros}($N$,1)
  \State $i = 1$  \% Set the current index
  \State $Ava\_State$ = \textbf{find}($T$(i,:)==1)
  \For{$j=1:$ \textbf{length}($Ava\_State$)}
    \State $d_1=0$
    \State [$Path, d_1, Cnt, P$]
    \State \;\;\;=\textbf{PathProp}($T, Path, d_1, d, i, Ava\_State(j), Cnt, P$)
  \EndFor
  \State Return $P$ and the indices set $\mathcal{D}$ of rows in which there are elements larger than $d+1$.
  \end{algorithmic}
\end{algorithm}

\begin{lem}\label{lem:upbd}
Suppose $P$ is the output of Algorithm \ref{Alg:alg2} for two aligned vector $X_1^N$ and $Y_1^{N-d}$, and $\hat{d}$ is the number of rows of $P$ with elements larger than $d+1$. The expectation of $\hat{d}$ satisfies $\mathbb{E}[\hat{d}] \leq 3d$.
\end{lem}
\begin{IEEEproof}
Consider $d=1$ firstly. The value of $\hat{d}$ gets larger than $1$ when the deletion occurs in more than one consecutive ``0"s or ``1''s. The probability of the deleted bit being covered by $n$ consecutive ``0''s or ``1''s is $\frac{n}{2^{n+1}}$.
\begin{eqnarray}
\mathbb{E}[\hat{d}]&=&\sum_{n=1}^{N} n\cdot \frac{n}{2^{n+1}} \\
&\leq& \sum_{n=1}^{\infty} n\cdot \frac{n}{2^{n+1}}\\
&=& 3.
\end{eqnarray}
Now consider the case when $d>1$. If all the deletions occur in different subsequences with consecutive ``0"s or ``1''s, the above inequality can be applied and we have $\mathbb{E}[\hat{d}] \leq 3d$. If two deletions occur in the same subsequence with consecutive ``0"s or ``1''s, the potential deletion indices overlap and $\mathbb{E}[\hat{d}]$ becomes smaller. Therefore, the upper bound $\mathbb{E}[\hat{d}] \leq 3d$ still holds.
\end{IEEEproof}

\begin{algorithm}[H]
  \caption{The path propagation algorithm for a given state}
  \label{Alg:alg3}
  \begin{algorithmic}[1]
  \State \textbf{Input}: the admissible table $T$, the current path vector $Path$, the previous number of deletions $d_1$, total number of deletions $d$, current index $i$, the chosen state $State$, counter of available paths $Cnt$, the potential path table $P$
  \State \textbf{Output}: the updated path vector $Path$, the previous number of deletions for the next index $Next\_d_1$, the updated counter of available paths $Cnt$, the updated potential path table $P$
  \State $N=$ \textbf{length}($Path$)
  \If{$i==N$}
    \If{$State<d+2$ and $(State-1)==d_1$}
      \State $Next\_d_1=d_1$
      \State $Path(i)=State$
      \State $P(:,Cnt+1)=Path$
      \State $Cnt=Cnt+1$
      \State \Return
    \ElsIf {$State>d+1$ and $(State-d-2)==d_1$}
      \State $Next\_d_1=d_1+1$
      \State $Path(j)=State$
      \State $P(:,Cnt+1)=Path$
      \State $Cnt=Cnt+1$
      \State \Return
    \Else
      \State $Next\_d_1=d_1$
      \State \Return
    \EndIf
  \Else
    \If{$State<d+2$ and $(State-1)==d_1$}
      \State $Next\_d_1=d_1$
      \State $Path(i)=State$
      \State $Ava\_State=$ \textbf{find}($T(i+1,:)==1$)
      \For{$j=1:$ \textbf{length}($Ava\_State$)}
        \State $[Path, NNext\_d_1, Cnt, P]$=
        \State \;\;\;\textbf{PathProp}($T, Path, Next\_d_1, d, i+1, ...$
        \State \;\;\;\;\;\;\;\;\;\;\;\;...$Ava\_State(j), Cnt, P$)
      \EndFor
    \ElsIf{$State>d+1$ and $(State-d-2)==d_1$}
      \State $Next\_d_1=d_1+1$
      \State $Path(i)=State$
      \State $Ava\_State=$ \textbf{find}($T(i+1,:)==1$)
      \For{$j=1:$ \textbf{length}($Ava\_State$)}
        \State $[Path, NNext\_d_1, Cnt, P]$=
        \State \;\;\;\textbf{PathProp}($T, Path, Next\_d_1, d, i+1, ...$
        \State \;\;\;\;\;\;\;\;\;\;\;\;...$Ava\_State(j), Cnt, P$)
      \EndFor
    \Else
      \State $Next\_d_1=d_1$
      \State \Return
    \EndIf
  \EndIf
  \end{algorithmic}
\end{algorithm}

The numerical simulation result of the relationship between $\mathbb{E}[\hat{d}]$ and $d$ for $N=256$ can be found in Table \ref{tab:dbound}, which shows that the upper bound in Lemma \ref{lem:upbd} is tight.

\begin{table}[ht]
\begin{center}
\caption{The relationship between $\mathbb{E}[\hat{d}]$ and $d$ for $N=256$}
\label{tab:dbound}
\begin{tabular}{|c||c|c|c|c|c|c|}\hline
 & $d=1$ & $d=2$ & $d=3$ & $d=4$ & $d=5$ & $d=6$ \\\hline\hline
$\mathbb{E}[\hat{d}]$  &  2.9985  &  5.9593  & 8.9893 & 11.9026 & 14.8974 & 17.7470  \\\hline
\end{tabular}
\end{center}
\end{table}

\begin{rem}
One may be curious about the number of potential deletions if more columns are aligned. Let $\hat{\hat{d}}$ denote the detected number of deletions after two-column alignment between Alice and Bob. More explicitly, let $\dot{X}_1^N$ ($\ddot{X}_1^N$) and $\dot{Y}_1^{N-d}$ ($\ddot{Y}_1^{N-d}$) denote the first (second) data column of Alice and Bob, respectively. We can invoke Algorithm \ref{Alg:alg2} for two times to obtain two sets of deletion indices $\mathcal{D}_1$ and $\mathcal{D}_2$. Clearly, the set of potential deletion indices can be shrinked to $\mathcal{D}=\mathcal{D}_1\cap \mathcal{D}_2$, and we have $\hat{\hat{d}} \leq \hat{d}$. Numerical result shows that $\mathbb{E}[\hat{\hat{d}}] \approx 1.7d$. The expectation can be further reduced to roughly $1.3d$ and $1.1d$ when three and four columns are used, respectively.
\end{rem}

\section{Feedback and Lossless Compression}

\begin{figure}[ht]
    \centering
    \includegraphics[width=9 cm]{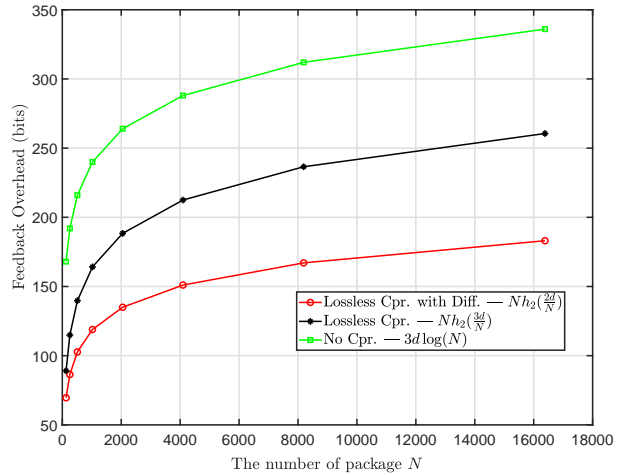}
    \caption{A comparison of the feedback overhead for different compression schemes with $d=4$ and $N=\{2^7,2^8,...,2^{14}\}$. The green curve (marked with squares) stands for directly sending the indices of the potentially missing packages, the black curve (marked with stars) represents the lossless compression scheme for the deletion state sequence $D_1^N$, and the red curve (marked with circle) labels the lossless compression scheme for the differential version of $D_1^N$.}
    \label{fig:CmprOrNot}
\end{figure}

After identifying the potential deletion positions, the rest of Bob's task is to feedback these positions to Alice, who is going to send the corresponding packages and complete the reconciliation process. A natural way is to send Alice the indices of the potentially missing packages. Since there are $N$ packages in total, each index can be represented by $n=\log N$ bits. Therefore, when a single column is used for the deletion detection, the average overhead for sharing the missing indices in roughly $3dn$. However, by taking the advantage of source polarization, we may reduce this overhead. Recall that the missing state can be represented by a bit for each package, e.g., 1 stands for missing and 0 stands for the opposite. Then, the overall states of the $N$ packages can be expressed by an $N$-bit sequence, denoted by $D_1^N$. As a result of Lemma \ref{lem:upbd}, the sequence $D_1^N$ is relatively biased, with roughly $3d$ ones and $N-3d$ zeros. We may simply treat $D$ as a Bernoulli random variable with distribution $Ber(p)$\footnote{It should be noted that the state random variable $D$ is not independently distributed. However, we may use a pre-shared random permutation between Alice and Bob to remove the dependency.}, where $p=\frac{3d}{N}$. Consequently, the entropy of $D$ is given by $ h_2(\frac{3d}{N})$, which means that $D_1^N$ can be represented by roughly $N\cdot h_2(\frac{3d}{N})$ bits. A comparison between these two overheads $3dn$ and $N\cdot h_2(\frac{3d}{N})$ for $d=4$ and various $N$ is shown in Fig. \ref{fig:CmprOrNot}. It can be seen that the overhead after lossless compression can be slightly improved, which explains our motivation to some extend.

By taking a closer look at $D_1^N$, one may find that the compression rate can be further reduced. By the analysis in the proof of Lemma \ref{lem:upbd}, the ambiguousness of the deletion positions is mainly caused by the consecutive ``0"s or ``1"s in the sequence $X_1^N$, which results in consecutive ``1"s in the sequence $D_1^N$. A differential operation \footnote{To maintain the length $N$, we assume a padding 0 at the beginning of the sequence $D_1^N$ before the differential operation.} on $D_1^N$ can break the segments of consecutive ``1"s and make the proportion of ``1''s smaller, which leads to a better compression rate. We have the following lemma.

\begin{lem}\label{lem:diffbnd}
Let $D_1^N$ denote the $N$-bit sequence labeling the state of deletion of each package after aligning $X_1^N$ and $Y_1^{N-d}$ according to Algorithm \ref{Alg:alg2}. Suppose $\bar{D}_1^N$ is the differential version of $D_1^N$, and $\bar{d}$ is the number of ``1"s in $\bar{D}_1^N$. The expectation of $\bar{d}$ satisfies $\mathbb{E}[\bar{d}] \leq 2d$.
\end{lem}
\begin{IEEEproof}
Similarly to the proof of Lemma \ref{lem:upbd}, we consider $d=1$ firstly. The probability of the deleted bit being covered by $n$ consecutive ``0"s or ``1"s is $\frac{n}{2^{n+1}}$. This event would result in $n$ consecutive ``1"s in $D_1^N$. After the differential operation, only 2 of them are left. Consequently, the expectation can be calculated as
\begin{eqnarray}
\mathbb{E}[\bar{d}]&=&\sum_{n=1}^{N} 2\cdot \frac{n}{2^{n+1}} \\
&\leq& \sum_{n=1}^{\infty} 2\cdot \frac{n}{2^{n+1}}\\
&=& 2.
\end{eqnarray}

For the case when $d>1$. We can similarly claim that $\mathbb{E}[\bar{d}] \leq 2d$, because some deletions may occur in a same segment of consecutive ``0"s or ``1"s, shrinking the number of ``1"s in $\bar{D}_1^N$.
\end{IEEEproof}

The numerical simulation result of $\mathbb{E}[\bar{d}]$ for various $d$ and $N=256$ is given in Table \ref{tab:dbarbound}, which shows that the upper bound in Lemma \ref{lem:diffbnd} is tight, especially for relatively small $d$ and large $N$. We then treat $\bar{D}$ as a Bernoulli random variable with distribution $Ber(\frac{2d}{N})$, whose entropy is given by $h_2(\frac{2d}{N})$ bits. The feedback overhead $N\cdot h_2(\frac{2d}{N})$ is also depicted in Fig. \ref{fig:CmprOrNot} for comparison.

\begin{table}[ht]
\begin{center}
\caption{The relationship between $\mathbb{E}[\bar{d}]$ and $d$ for $N=256$}
\label{tab:dbarbound}
\begin{tabular}{|c||c|c|c|c|c|c|}\hline
 & $d=1$ & $d=2$ & $d=3$ & $d=4$ & $d=5$ & $d=6$ \\\hline\hline
$\mathbb{E}[\bar{d}]$  &  1.9927  &  3.9389  & 5.8482 & 7.6799  & 9.4958 & 11.2399  \\\hline
\end{tabular}
\end{center}
\end{table}

According to Shannon's source coding theorem, the average compression rate can be made arbitrarily close to the source entropy, i.e., the compression rate $h_2(\frac{2d}{N})$ can be asymptotically achieved for the source $\bar{D}$. Thanks to the technique of source polarization \cite{polarsource,cronie2010lossless}, we may still use polar codes to complete this task. With some abuse of notation, let $U_1^N=\bar{D}_1^N G_N$ denote the sequence after the polar transform. The source polarization theorem says that as $N$ grows, almost all the conditional entropy $H(U_i|U_1^{i-1})$ for $i \in [N]$ polarizes to 0 or 1. Moreover, the proportion of the indices with $H(U_i|U_1^{i-1})$ close to 1 approaches to $H(\bar{D})$, and those with $H(U_i|U_1^{i-1})$ close to 0 approaches to $1-H(\bar{D})$. Let $\mathcal{S}$ denote the subset of $[N]$ such that $H(U_i|U_1^{i-1}) \to 0$ for $i \in \mathcal{S}$. Then, the source sequence $\bar{D}_1^N$ can be compressed into the subsequence of $U_1^N$ with indices in $\mathcal{S}^c$, which is denoted by $U^{\mathcal{S}^c}$. For recovery, since $H(U_i|U_1^{i-1}) \to 0$ for $i \in \mathcal{S}$, the bits with indices in $\mathcal{S}$ can be decoded from $U^{\mathcal{S}^c}$ with vanishing error probability by using standard decoding algorithms of polar codes. To guarantee a zero error probability for lossless compression, the source encoder can run the decoding algorithm and check if the estimate $\hat{U}^{\mathcal{S}}$ of $U^{\mathcal{S}}$ matches or not. Let $\mathcal{T}$ denote the subset of $\mathcal{S}$ such that $\hat{U}_i \neq U_i$ for $i \in \mathcal{T}$ by the decoding algorithm. The encoder sends $U_M=\{U^{\mathcal{S}^c}, \mathcal{T}\}$ to make sure that $U^{\mathcal{S}}$ can be correctly recovered at the side of decoder. Finally, $\bar{D}_1^N$ is reconstructed by $\bar{D}_1^N=U_1^N G_N^{-1}$ and in fact $G_N^{-1}=G_N$. We note that the proportion $\frac{|\mathcal{T}|}{N}$ tends to 0 for sufficiently large $N$.

For simplicity, we choose the standard SC decoding method for numerical simulation. The comparison of the average feedback overhead in bits between the direct feedback scheme and the compression scheme with differential operation is shown in Table \ref{tab:cpr}, where the overhead of direct feedback is given by $n \cdot \mathbb{E}[\hat{d}]$ and that of lossless compression is calculated by $|\mathcal{S}^c|+n\cdot\mathbb{E}[|\mathcal{T}|]$. It demonstrates that the feedback overhead can be further reduced, by the simple lossless compression scheme with complexity of $O(N\log N)$. We note that the compression rate can be further improved by using more sophisticated decoding algorithms \cite{Eslami2012BPfinite,ListPolar,NiuKaiSCS2013}. After recovering $\bar{D}_1^N$ and then $D_1^N$, Alice sends the corresponding packages to Bob, which completes the reconciliation process.

\begin{rem}
It is possible to use the network coding technique \cite{AhlswedeNetwork2000} to reduce the number of sending packages on Alice's side, because the genuine number $d$ of deletions is no larger than $\hat{d}$. An intuitive example is the case when $d=1$, and $\hat{d} \geq 1$ can be any integer. When Alice recovers $D_1^N$ and locates the $\hat{d}$ potential deletions successfully, she does not need to send the $\hat{d}$ corresponding packages to Bob. Instead, sending a single ``checksum" package of all the $\hat{d}$ packages to Bob is sufficient to help Bob recover the missing package. When $d>1$, how to design the network coding scheme to optimize the number of sending packages on Alice's side is a future work.
\end{rem}

\begin{table}[ht]
\begin{center}
\caption{Numerical simulation of the average overhead (in bits) for various $d$ and $N$.}
\label{tab:cpr}
\begin{tabular}{|c||c|c|}\hline
  & Direct Feedback &  Cpr. with. Diff. \\\hline\hline
  $d=8$,  $N=256$&    189.8272  &   101.2584    \\\hline
  $d=10$, $N=256$&    237.7520  &   114.1440    \\\hline
  $d=8$, $N=512$&     215.6850  &   131.4060    \\\hline
  $d=10$, $N=512$&    266.5980  &   150.6010    \\\hline
  $d=8$,  $N=1024$&   239.2500  &   161.7400    \\\hline
  $d=10$, $N=1024$&   293.3600  &   188.2800    \\\hline
  $d=20$, $N=1024$&   582.9000  &   306.0160    \\\hline
\end{tabular}
\end{center}
\end{table}

\section{Conclusion}
In this paper, we proposed a total polar coding based set reconciliation scheme between two network nodes which are sharing data with unknown deletions. Firstly a polar code aiming to help one node to recover a certain amount of the other's data is constructed in the presence of deletions. The problem is modeled as the Slepian-Wolf coding with deletions, which can be solved by designing polar codes for deletion channels. By aligning the local data with the recovered data of the other, the position of potential deletions can be revealed. We also designed an explicit algorithm for this aligning process. After that, a lossless compression scheme based on source polarization is utilized to reduce the feedback overhead of the deletion position information as much as possible. Our scheme is immune to the size of the data package, and the overall complexity is only related to the package number $N$, which is particularly given by $O(N \log N)$ if the number $d$ of deletions is fixed.  We also provided some analysis on the upper bound of the number of detected deletions.





%
\bibliographystyle{IEEEtran}
\bibliography{Myreff}

\end{document}